\documentclass[12pt]{iopart}
\usepackage{graphicx}
 
\begin{document}

\title{Optical black hole lasers}

\author{Daniele Faccio$^{1}$, Tal Arane$^{1}$, Marco Lamperti$^{1}$, Ulf Leonhardt$^{2}$}

\address{$^{1}$School of Engineering and Physical Sciences, SUPA, Heriot-Watt University, EH14 7JG, Edinburgh, UK.\\
$^{2}$School of Physics and Astronomy, University of St Andrews, North Haugh, St Andrews KY16 9SS, UK.}
\ead{d.faccio@hw.ac.uk}
\begin{abstract}
Using numerical simulations we show how to realise an optical black hole laser, i.e. an amplifier formed by travelling refractive index perturbations arranged so as to trap light between a white and a black hole horizon. The simulations highlight the main features of these lasers: the growth inside the cavity of positive and negative frequency modes accompanied by a weaker emission of modes that occurs in periodic bursts corresponding to the cavity round trips of the trapped modes. We then highlight a new regime in which the trapped mode spectra broaden until the zero-frequency points on the dispersion curve are reached. Amplification at the horizon is highest for zero-frequencies, therefore leading to a strong modification of the structure of the trapped light. For sufficiently long propagation times, lasing ensues only at the zero-frequency modes. 

\end{abstract}


\maketitle

\section{Introduction}
Since Unruh's first suggestion that flowing media can be used as analogues for gravity \cite{unruh} and in particular for studying certain physical phenomena usually associated to gravitational event horizons, e.g. Hawking radiation, the field has seen a steady increase in the number of proposed physical systems in which to observe such effects \cite{novello_book,barcelo_review}. Originally, analogue gravity was proposed in systems that physically display a flowing medium that reproduces the flow of space close to a black hole: a flowing fluid that exhibits a gradient from subsonic to supersonic flow presents a point within the gradient such that the flow speed equals the speed of acoustic waves. This is a point of non-return for acoustic waves trying to propagate against the flow and is the analogue of an event horizon, causally disconnecting two separate regions within the medium. Such horizons are much more than mere toys or superficial analogies: they can truly reproduce the kinematics of gravitational event horizons. This implies that a direct analogue of the amplification of vacuum fluctuations and consequent emission of particles away from the horizon, known as Hawking radiation, should be visible. This idea has of course attracted significant attention along with the proposal of various systems that may enhance the Hawking emission \cite{barcelo_review}. The main point is that Hawking emission is predicted to occur with a blackbody distribution with a temperature $T$ connected to the horizon through the so-called surface gravity, $\kappa$, i.e. the gradient of the gravitational field across the horizon \cite{hawking1,hawking}. In analogue systems, the surface gravity is determined by the gradient of the medium flow across the horizon and the combination of steep gradients along with very cold environments (in order to reduce the background thermal noise) is expected to lead to the observation of the spontaneous Hawking emission. Recently Philbin et al. proposed an optical analogue for horizons in which the steep gradients are ensured by shock front formation in intense optical pulses and the reduction of the background thermal noise is ensured by the fact that at optical frequencies, room temperature thermal noise is completely negligible \cite{philbin}. \\
In analogue models,  black holes and their time-reversed realisation, i.e. white holes \cite{hawking_WH} play equivalently important roles. Whilst gravitational white holes are expected to be extremely rare, or even non-existent, analogue models abound with ways to generate such objects. Most notably, recent experiments performed in the presence of a white hole horizon in flowing water highlighted a stimulated frequency conversion process that can be considered as the classical limit of Hawking emission \cite{rousseaux,weinfurtner} whilst an optically induced white hole horizon was used to observe both the classical shifting of light frequency \cite{philbin,faccio,transistor,konig} and the spontaneous emission of photons \cite{belgiorno,rubino} that has stimulated an ongoing discussion regarding the precise origin of the emission \cite{comment1,reply1,liberati2011}.\\
A remarkable prediction that emerged from analogue gravity studies was the prediction that a combination of a black hole and white hole event horizon would lead to a feed-back amplification process very similar to a laser where radiation is continuously bounced back and forth between the two  horizons and hence exponentially amplified \cite{laser}.  \\
Black hole lasers were first proposed proposed for systems that exhibit a ``superluminal'' dispersion relation (i.e. in which the group velocity increases with frequency) such as, for example, Bose-Einstein condensates (BECs). Black hole lasing has subsequently been studied both theoretically and numerically in Bose-Einstein-Condensate systems and similar settings \cite{laser,ulf-laser,parentani-laser,finazzi-laser}. In this work we present a numerical study of black hole lasing in optical systems. Our numerics account for realistic parameters of the medium  and highlight a laser amplification mechanism that is similar to that studied in BECs. However, we also show that as evolution ensues, the laser develops low frequency components that eventually seed amplification at what we call the ``zero-frequency'' modes. This amplification regime is exponentially favoured over the traditional black hole laser regime and leads to a significant increase of the output radiation. We then highlight and study how the low (close to zero) frequency region of the spectrum determines the properties of the high frequency emission, a feature that is expected to be generic to all analogue horizon systems.\\
\section{Optical horizon analogues.}
Optically induced horizon analogues are based on the use of nonlinear optics to create an effective moving medium \cite{philbin,rubino,faccioCP}. An intense laser pulse with carrier frequency $\omega_p$, typically in the near-infrared or visible region, is focused into a dielectric medium and through the nonlinear Kerr effect it will excite a material polarisation response that contains two terms. One term oscillates at frequency $2\omega_p$ and may excite photons from the vacuum state under the condition that energy and momentum conservation relations are satisfied. Unless particular care is taken in order to enforce such conditions, this so-called four-wave-mixing term (four elementary excitations are involved, two from the input pulse and two from the vacuum state) will be very strongly suppressed \cite{boyd}. There is a second term that follows the envelope of the input pulse, i.e. it responds to the DC component of the electromagnetic field and is described in terms of a variation of the medium refractive index, $n=n_0+n_2I(\zeta)$, where $n_0$ is the background refractive index ($n_0\sim1.45$ in most glasses) and $n_2$ is the nonlinear refractive, also called ``Kerr'' index of the medium and $\zeta=z-vt$ is the local lognitudinal cordinate. $I(\zeta)$ is the propagating intensity (envelope) profile of the laser pulse. In most common media $n_2\sim 10^{-16}$ W/cm$^2$ so that the maximum index variation, $\delta n_{\textrm{max}}\sim0.005-0.0001$, depending on the laser pulse intensity. With other media it may be possible to obtain higher $\delta n$, e.g. some so-called soft glasses or media such as graphene are expected to give values larger by an order of magnitude or more.\\
Light travels slower in high refractive index regions and faster in low refractive regions, so the moving laser pulse creates a moving $\delta n$ that locally slows light down. The effective space-time metric associated to the moving $\delta n$ is the so-called Gordon metric \cite{gordon,cacciatori} and may be recast in a form that tightly resembles the Painlev\'e-Gullstrand metric for black holes \cite{pain,gullstrand,belgiorno}. In other words, in the reference frame comoving with the $\delta n$, space is flowing with velocity $V=\gamma^2 v (n^2-1)/n$: on the leading edge of the perturbation space flows inwards with increasing velocity until $V=c$ (i.e. $v=c/n$) and a black hole horizon is formed. On the trailing edge, the time reversed situation is verified with space flowing outwards at a decreasing velocity and when $V=c$ a white hole horizon is formed (see also \cite{faccioCP}). \\
In the presence of dispersion, i.e. of a frequency dependence of the medium background refractive index, one must take care in defining the exact nature of the horizon due to the fact that the phase and group velocities differ. In the following we will always consider the case in which the horizon is a blocking point for the {\emph{group}} velocity of light, in line with the generally accepted opinion that this kind of horizon is the analogue for photons (or waves in a flowing medium) of a gravitational event horizon.\\
\begin{figure}
\begin{centering}
\includegraphics[width=15cm]{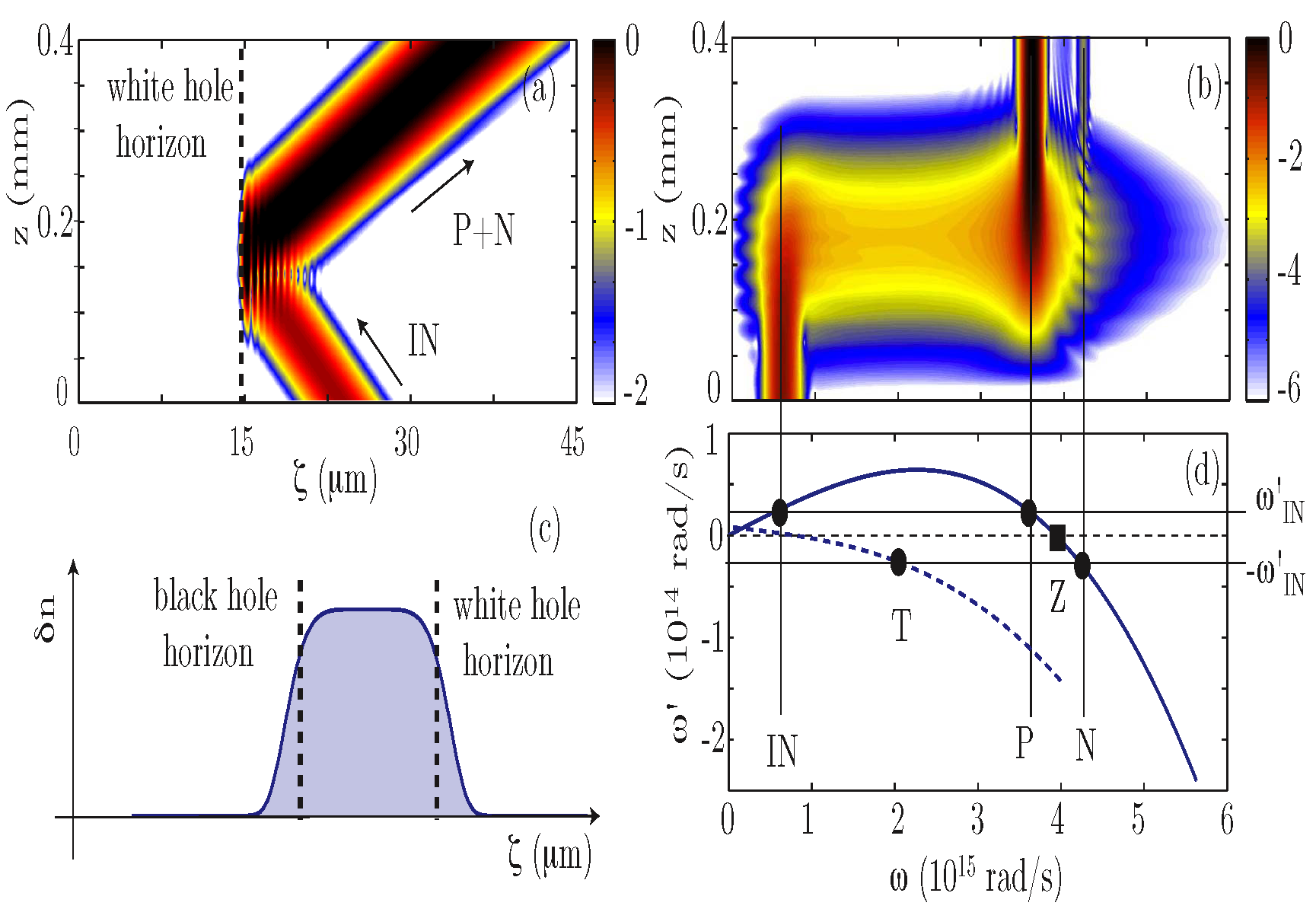}
\caption{Numerical simulations of the interaction of a laser pulse with a white hole horizon. (a) Amplitude  profile, in logarithmic scale, of the laser pulse reflecting on the horizon (dashed line). (b) Evolution (along the propagation direction, z) of the spectrum, in logarithmic scale, during reflection from the horizon. (c) Schematic overview of the the geometry of a generic refractive index perturbation used in this work and location of the black and white hole horizons. (d) Dispersion curve used in this work with indicated the positions of the IN (input), P (positive), N (negative), Z (zero-frequency) and T (transmitted) modes. All modes are determined by the condition that frequency is conserved in the comoving frame, i.e. by the intersections of horizontal lines passing through the IN mode frequency, $\pm\omega_{\textrm{IN}}$.}
\label{horizons}
\end{centering}
\end{figure}

\section{Horizons in dispersive media.}
\begin{centering}
\begin{figure}
\begin{centering}
\includegraphics[width=15cm]{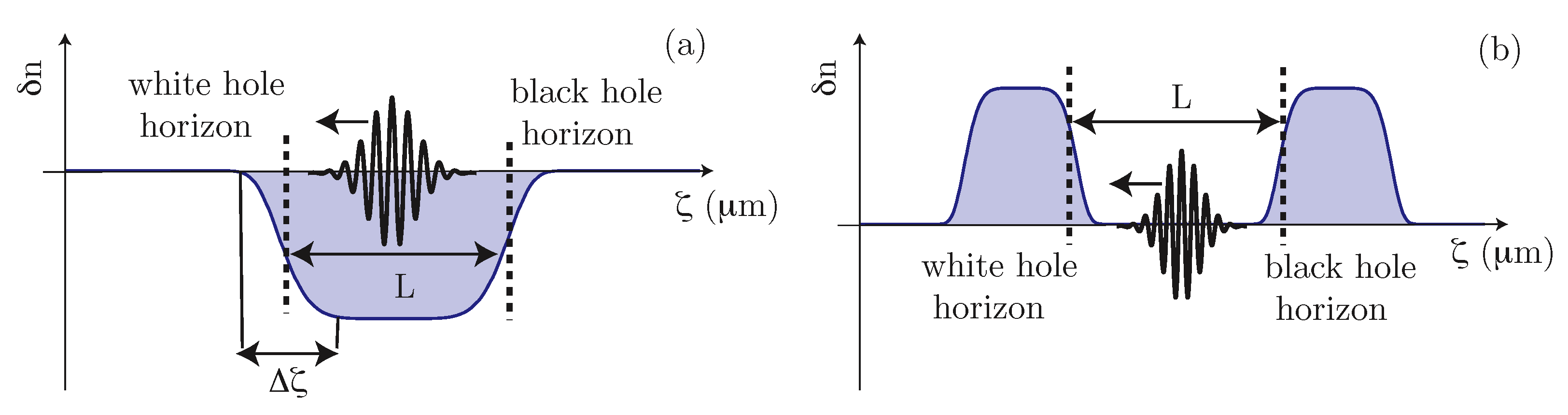}
\caption{Possible configurations of the optical black hole laser in the presence of subluminal dispersion. The seed laser pulse is injected directly in the cavity and then propagates first towards the white hole horizon (as indicated by the arrows).}
\label{laser}
\end{centering}
\end{figure}
\end{centering}

We now study how an input wave behaves when interacting with such horizons. This behaviour is analysed by performing numerical simulations based on Maxwell's equations, discretized and solved following the so-called Pseudo-Spectral-Space-Domain (PSSD) technique \cite{pssd}  that is a variation of the more common FDTD technique \cite{fdtd_book} and allows to include arbitrary material dispersion in a very simple manner. In more detail, we consider 1-dimensional propagation modelled by the Maxwell equations for the transverse $x$ and $y$ components of the electric ($E_x$) and magnetic fields ($H_y$), respectively:
\begin{eqnarray}
\partial_z E_x &=&-\mu_0\partial_tH_y\\
\partial_z H_y &=&-\partial_tD_x
\end{eqnarray}
The electric displacement field is numerically evaluated as $D_x(t)=\mathcal{F}^{-1}\{\varepsilon(\omega)\mathcal{F}[(1+\varepsilon(z-vt))E_x(t)]\}$, where $\mathcal{F}$ and $\mathcal{F}^{-1}$ indicate the Fourier and inverse Fourier transforms, $\varepsilon(\omega)$ is the dispersive part of the medium refractive index, $n_0=\sqrt{\varepsilon}$. The moving refractive perturbation  is described by a super-gaussian function $\varepsilon(z-vt)=\varepsilon_0\exp[-(z-vt)^{2m}/\sigma^{2m}]$, where $m$ is the super-gaussian order and controls the perturbation steepness and $\sigma$ gives the perturbation width, typically fixed at $50$ $\mu$m. The black hole cavity is constructed by using two such perturbations, displaced along the $\zeta$-axis by a distance $L$ that fixes the cavity length (see Fig.~\ref{laser}). The input seed laser pulse is made to start directly inside the cavity and has the functional form $E(z-vt)=\exp[-(z-vt)^{2}/\sigma_E^{2}]\cos(kz-\omega t)$, where the initial pulse width $\sigma_E$ is always taken to be shorter than the cavity length $L$, e.g. $\sigma_E\sim3-6$ $\mu$m and $L\sim 20-80$ $\mu$m.  The photon numbers generated at each frequency are evaluated as $S(\omega,z)/\omega$, where $S(\omega,z)$ is the spectral intensity evaluated at each propagation distance, $z$. Total output photon numbers are obtained by integrating this quantity over frequency. The code was implemented with a dispersion relation shown by the solid line in Figure~\ref{horizons}(d) in $(\omega^{\prime},\omega)$ coordinates, where $\omega^{\prime}=\omega-vk$ with $k=(\omega/c)n(\omega)$, is the frequency in the comoving reference frame. As can be seen, we are using a simplified dispersion curve that does not present any of the typical resonances present in most media such as glass. However, it was chosen to qualitatively reproduce the dispersion of diamond, a material that is currently becoming rather popular for a variety of reasons, including its remarkable transparency range due precisely to the absence of material resonances from sub-THz wavelengths up to the near-UV. Most importantly, such a choice allows us to study the generic physics underlying the black hole laser without additional complications due to specific structures within the dispersion relation. \\
 Figures~\ref{horizons}(a) and (b) illustrate the behaviour at a white hole horizon: the input mode IN is chosen with a low frequency (3 $\mu$m wavelength) such that it travels with positive group velocity in the comoving frame (the IN mode identified on the dispersion curve in (d) has positive gradient) and therefore catches up with the travelling $\delta n$. Upon reaching the horizon, light is blocked and transformed in to two new modes, one with positive frequency, P, and one with negative frequency, N. These both have negative group velocity in the comoving frame and are therefore reflected away from the 
 horizon (see (a)). These two modes are the Hawking modes generated at the horizon. Most importantly we have explicitly verified that, in agreement with the interpretation in terms of Hawking emission, these two modes are such that the difference of their squared norms (i.e. their photon numbers) is equal to one, $|P|^2-|N|^2=1$ and the ratio of the squared norms decays exponentially with increasing comoving frequency, $|N|^2/|P|^2\propto \exp(-C\omega^{\prime})$, where $C=[(2\pi c)/(\gamma^2v^2)]1/(d\delta n/d\zeta)$ is the decay constant  and $d\delta n/d\zeta$ is the perturbation gradient evaluated at the horizon \cite{belgiorno,rubino}. These two relations imply that the horizon emission in the P-mode exhibits a planckian black-body spectrum as a function of frequency. Most importantly for the black hole laser, the first relation also implies that the total photon number in reflection from the white hole horizon $|P|^2+|N|^2>1$, i.e. the white hole horizon acts as an amplifying mirror. %The decay constant $C$ is determined by the steepness of the refractive index perturbation, i.e. $1/C\propto dn/d\zeta$ where the gradient is evaluated at the horizon. 
 For small frequencies the photon gain $|P|^2+|N|^2$ scales as $1/C\omega^{\prime}$, i.e. the steeper the gradient the larger the horizon gain will be. This simple recipe is important in order to understand and optimise black hole lasers.  \\
 \begin{centering}
 \begin{figure}
\begin{centering}
\includegraphics[width=15cm]{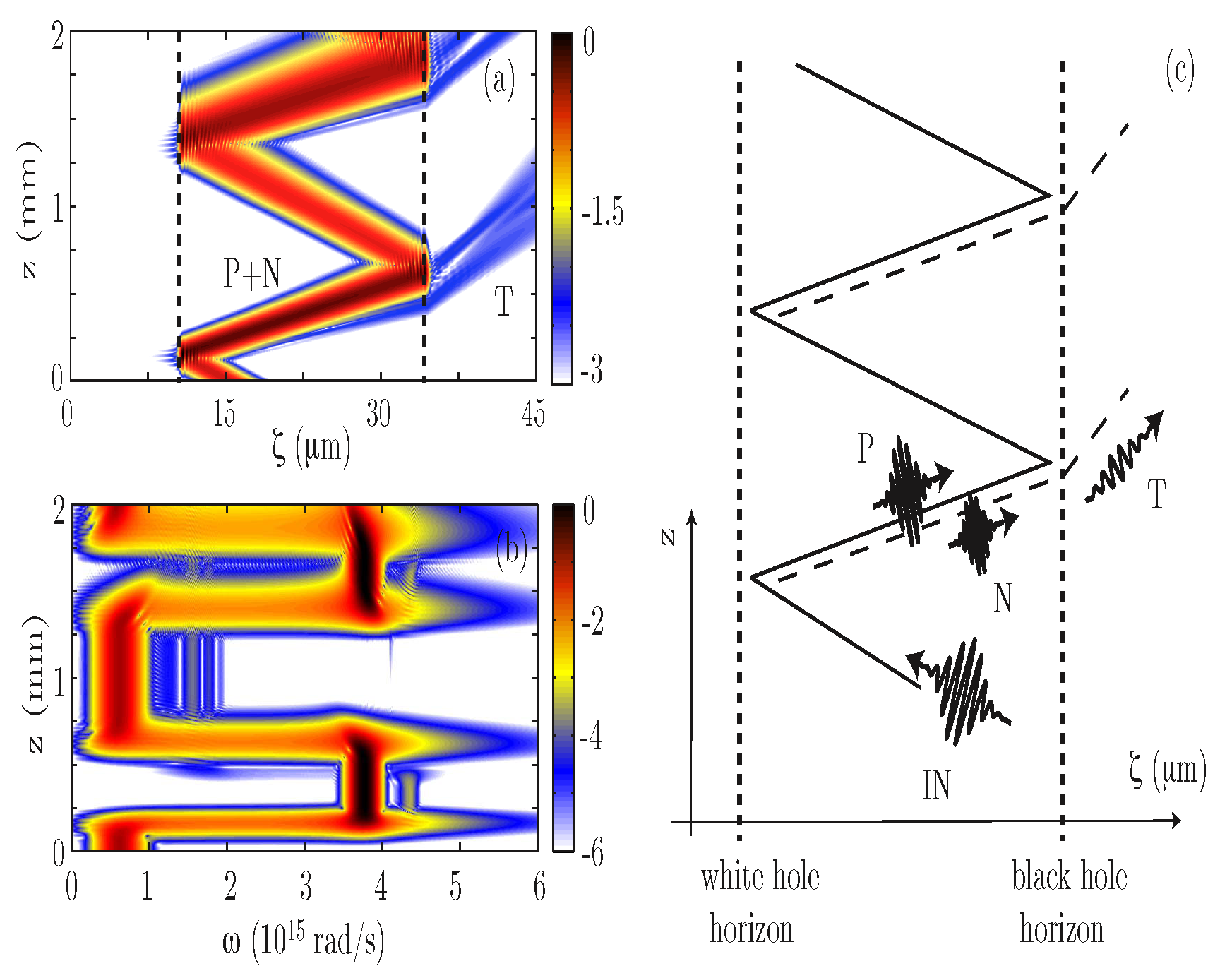}
\caption{Evolution of a laser pulse during the first few bounces from a white-black hole cavity (logarithmic scale). (a) Evolution along the propagation direction, $z$, of the amplitude in the comoving frame as the pulse bounces back and forth. (b) Evolution of the pulse spectrum. (c) Schematic representation of the cavity behaviour and indication of the various modes that are produced during each round-trip.}
\label{3bounces}
\end{centering}
\end{figure}
\end{centering}
The behaviour of the black hole horizon can be inferred from that of the white hole  as one is the time-inverted version of the other. This implies that a single mode impinging on the black hole horizon will not lead to negative mode generation or more in general to positive/negative mode mixing. The P mode will simply convert to the IN mode and reflect away from the black hole horizon. The N mode on the other hand will convert to a second red-shifted mode that still has negative group velocity and is thus transmitted through the horizon. This transmitted mode is indicated with ``T'' in Fig.~\ref{horizons}(d) (the dashed dispersion curve is relative to the region inside the black hole). \\
 Mode mixing and amplification at the black hole horizon will occur if both positive and negative modes, e.g. the same modes generated at the white hole horizon shown in Fig.~\ref{horizons}(a), are sent on to the horizon as initial input conditions.
%is somewhat more complicated and is illustrated in Figures~\ref{horizons}(c) to (h). In Figs.~\ref{horizons}(c) and (b) we first consider the case in which the P-mode alone encounters a black hole horizon. The mode travels slower than the $\delta n$ and is thus caught up by the black hole horizon where it is frequency converted to the red-shifted IN mode and thus reflected away due to the positive N-mode group velocity. We note that a black hole horizon approached from the left will only red-shift the mode frequency whilst the negative modes on the dispersion lie at higher frequencies. Therefore negative frequencies will not be generated in this process and amplification will not occur.\\
%The negative N-mode suffers a different fate as shown in Figures~\ref{horizons}(e) and (f): the mode tunnels through the horizon and is frequency converted to a mode that now lies on the dispersion curve relative to the inside region of the $\delta n$, i.e. $k=(\omega/c)(n(\omega)+\delta n)$ (dashed line in Fig.~\ref{horizons}(g)). This mode indicated with ``T'' can be seen to still have negative group velocity and is therefore sucked in to the black hole as seen in Fig.~\ref{horizons}(e).\\
%A different scenario occurs when both the positive and negative modes are simultaneously present at the black hole horizon, Figs.~\ref{horizons}(g) and (h). 
In this case we indeed have the time reversed situation of the white hole horizon. However, as pointed out in \cite{ulf-laser}, this amplification now relies critically on the phases between the P and N modes. For the case of large bandwidth pulses such as those used here, some frequency components will have the correct phase for amplification whilst others will have the wrong phase. Moreover, it is extremely difficult to purposely control these phases due to the varying and relatively strong dispersion (i.e.  accumulated phases in traversing the cavity) across the pulse spectrum. We find that on average the P and N modes meeting at the black hole horizon lead to amplification, although this amplification is typically smaller than that occurring at the white hole horizon.\\
In the black hole laser configuration discussed below, the white hole horizon will be approached by a single IN mode whilst the black hole horizon will always be approached with a simultaneous combination of both P and N modes, i.e. the modes generated at the white hole horizon. Therefore, both horizons will act as amplifying mirrors.\\

\section{The optical black hole laser}
On the basis of this reasoning we can construct an intuitive picture of the behaviour of a black hole laser. 
The black hole laser is formed by placing a black hole horizon and a white hole horizon in close vicinity such that waves are trapped between the two. In order for this to happen care must be taken in choosing the correct arrangement and this depends on the dispersion. In previous studies based on phonon oscillations in Bose-Einstein-Condensates, a so-called superluminal dispersion was considered, i.e. the dispersion curve had positive curvature \cite{laser,ulf-laser,parentani-laser,finazzi-laser}. In our case we have a dispersion curve with a negative curvature and so-called subluminal dispersion. This implies that in the laboratory reference frame, the group velocity of light is always smaller than the phase velocity and also that it decreases with increasing frequency. Therefore,  based on the results in Fig.~\ref{horizons}, if we want to trap light between two horizons then the white hole horizon must be placed before (at smaller $\zeta$) the black hole horizon. \\
We illustrate this situation in Fig.~\ref{3bounces} that shows a numerical simulation of the first few cavity round-trips of a laser pulse trapped between two horizons. The spectrum evolves periodically at each reflection from the horizon mirrors. We clearly observe the same P and N modes seen in Fig.~\ref{horizons} at the white hole horizon. At the black hole horizon the spectrum returns to the original input frequency with the addition a second very weak, less red-shifted mode around $\omega=1.5\times10^{15}$ rad/s. This is the ``T'' mode mentioned in the previous section [see also Fig.~\ref{horizons}(d)]  that originates from the partial transmission of the N mode through the black hole horizon and is  ejected from the cavity. This mode is also visible in Fig.~\ref{3bounces}(a) although due to its very low intensity it does not play an important role in the laser dynamics reported in this work. A schematic representation of all the laser modes is shown in Fig.~\ref{3bounces}(c) that summarises the behaviour of the light pulse within the cavity and the interaction with the two horizon mirrors.\\
We note that it is possible to  achieve a black hole laser in two ways as shown in Fig.~\ref{laser}: (a) we may use a negative $\delta n$, similar to a potential well which will trap light. This is somewhat difficult to achieve of we assume that the $\delta n$ is generated by a laser pulse through the Kerr effect due to the fact that most media have positive $n_2$. Alternatively, (b) we may obtain the same effect by using two positive-valued perturbations generated by two independent laser pulses that are placed a certain distance apart.  \\
%We note that in the following we will always be considering a somewhat different situation with respect to previous studies, i.e. light is injected directly inside the cavity. Moreover,  many results of the this work refer to the case in which the IN mode is taken to be in the form a short pulse that is much shorter that the separation between the horizons, i.e. shorter than the cavity length (as schematically shown in Fig.~\ref{laser}). The pulse will therefore interact independently first with one horizon, be reflected and then interact with the other and therefore bounce back and forth, with amplification occurring upon each reflection. The longitudinal mode structure of the cavity can therefore be neglected  and will not lead to any selection of the propagating optical modes. Most importantly, the modes emitted from one horizon do not fill the whole cavity but only enter in contact with the opposite horizon after some propagation. Conversely, in previous studies the two horizons where directly ``connected'' by the modes that completely filled the cavity. 
%
\begin{centering}
\begin{figure}[t]
\begin{centering}
\includegraphics[width=15cm]{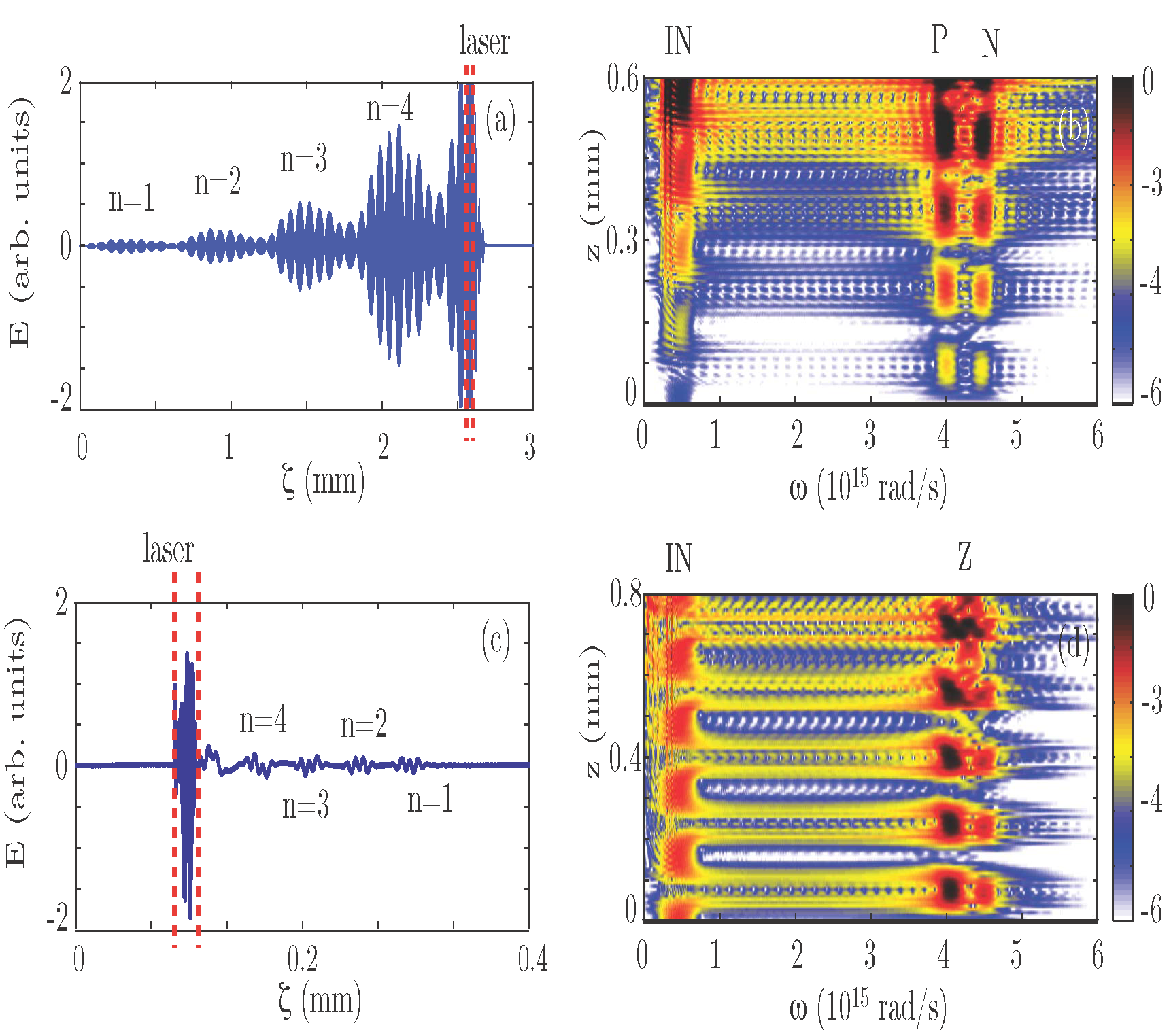}
\caption{Evolution of the laser spectrum (in log. scale) and electric field amplitude at the simulation output are shown for two different cases: (a)-(b) $\Delta\zeta=0.3$ $\mu$m, and (c)-(d) $\Delta\zeta= 3$ $\mu$m. All other parameters are identical for both simulations:  input wavelength 2 $\mu$m, cavity length L=20 $\mu$m, $\delta n_{\textrm{max}}=0.1$.}
\label{short_laser}
\end{centering}
\end{figure}
\end{centering}
\begin{centering}
\begin{figure}
\begin{centering}
\includegraphics[width=15cm]{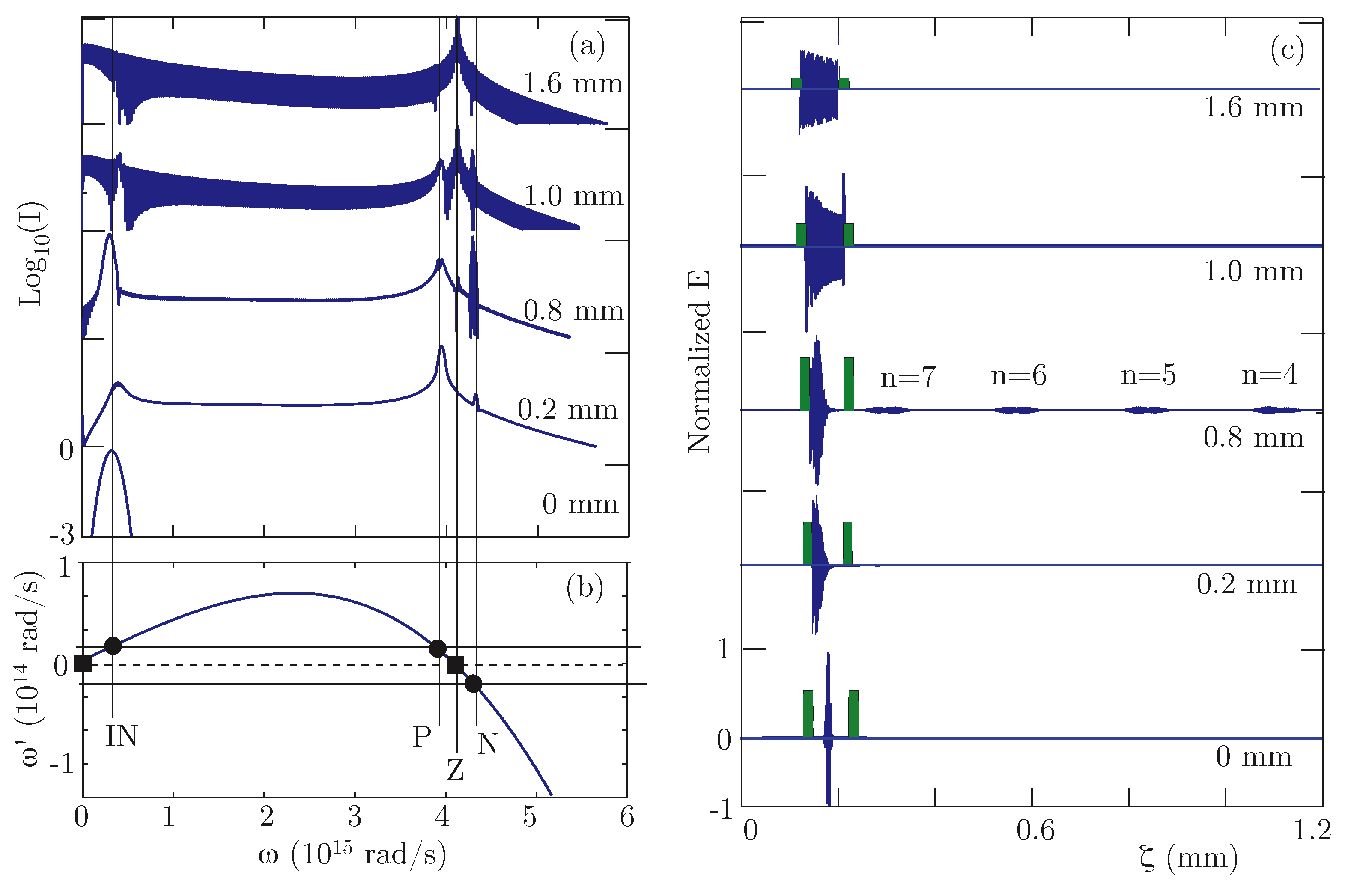}
\caption{Evolution of the normalised laser spectrum (in logarithmic scale, 3 decades) and normalised electric field amplitude (linear scale, from -1 to +1) - profiles are shown for increasing propagation distances as indicated in the graphs. Each 0.1 mm of propagation corresponds very closely (in this simulation) to one cavity round-trip. Parameters used in the numerical simulations are: input wavelength 2 $\mu$m, $\Delta\zeta=3$ $\mu$m, cavity length L=80 $\mu$m, $\delta n_{\textrm{max}}=0.01$. The shaded regions in (c) indicate the position and (rescaled) shape of the two $\delta n$ perturbations forming the cavity.}
\label{laser-sim}
\end{centering}
\end{figure}
\end{centering}
As an illustrative case, we show in Fig.~\ref{short_laser} the numerical results for a black hole laser that is obtained using a negative $\delta n$ ($\delta n_{\textrm{max}}=-0.1$) in which  the black and white hole horizons are separated by 4.5 $\mu$m. The input mode is a Gaussian shaped pulse, with 3 $\mu$m carrier wavelength and pulse length of 2 cycles.  Figure~\ref{short_laser}(a) shows the electric field in the vicinity of the black hole laser after a propagation distance of 650 $\mu$m. The vertical dashed lines indicated the position of the cavity that is travelling from left to right. The input mode, originally localised only within the cavity has led to lasing in the form of bursts of pulses tunnelling and exiting through the white hole horizon (not to be confused with the T-mode that would exit the black-hole horizon yet is too weak to be seen in this simulation). These bursts are composed of a sum of both P and N modes (hence the beat signal that modulates the individual bursts) and occur once every cavity round-trip, indicated with n=1,2,3,4 in the figure. These bursts are strongly increasing in amplitude at each round trip, a clear indication of the gain provided by the black hole laser. We note that here the modes leaking out of the cavity are doing so as a result of a tunnelling effect: the refractive index increase from the background to the maximum value (denoted by $\Delta \zeta$ in Fig.~\ref{laser}(a)) at the white hole horizon occurs over a distance that is shorter than the optical wavelength. The modes therefore tunnel out of the hole - this tunnelling may be controlled by reducing the gradient of the refractive index profile and indeed, for refractive index changes that occur over scales larger than the mode wavelength, the outgoing leakage is suppressed and the modes are completely trapped within the cavity (see below). Figure~\ref{short_laser}(b) shows the spectral evolution corresponding to Fig.~\ref{short_laser}(a): the periodic back and forth conversion between in the IN  and P/N modes at each cavity round trip can be clearly seen with a gain of more than 5 decades after only n=4 round trips. These findings are qualitatively similar to those reported for black hole lasers in Bose-Einstein condensates \cite{ulf-laser}. However, such a setting will be experimentally challenging both in virtue of the large $\delta n$ (although similar results were obtained with smaller $\delta n$,  at the cost of correspondingly longer propagation distances) and of the extremely short distances at play, i.e. the short separation between the horizons and sharp $\Delta\zeta$ increase at the horizon from the background index to the maximum $\delta n$. \\
In Figs.~\ref{short_laser}(c) and (d) we show the same simulation with a transition $\Delta \zeta$ that has been reduced to experimentally reasonable values, i.e. $\Delta \zeta \sim 3$ $\mu$m. As can be seen, the high frequency modes no longer leak out of the white hole horizon. This allows for a higher gain within the cavity itself, although this is largely counterbalanced by the detrimental effect of reducing the $\delta n$ gradient as this reduces the amplification at each reflection. Nevertheless, in this situation a stronger build-up of negative frequencies occurs inside the cavity, which partly leak out from the black hole horizon at each round trip in the form of a more intense and now visible ``T'' mode, as discussed in Fig.~\ref{horizons}(d). The T mode is emitted in bursts corresponding to the cavity round-trips (indicated with n=1,2,3,4 in the figure).\\
Finally, we note that in the very last stages of this simulation we can clearly see that the spectrum is starting to exhibit two distinct and new features: a strongly red-shifted component close to $\omega=0$ and a component, indicated with Z that lies in between the P and N modes.
 In the following, we focus attention on these new modes and highlight a new mechanism by which the pulse spectrum is broadened at each bounce until far-infrared wavelengths are generated. These then seed and sustain amplification at the zero-frequency points of the dispersion curve (indicated with filled squares in Fig.~\ref{horizons}(d)), which have a gain favoured by the aforementioned $1/\omega^{\prime}$ dependence. This results in remarkably stronger gain and significant lasing even in conditions that could be realised in experiments.\\
\begin{centering}
\begin{figure}
\begin{centering}
\includegraphics[width=15cm]{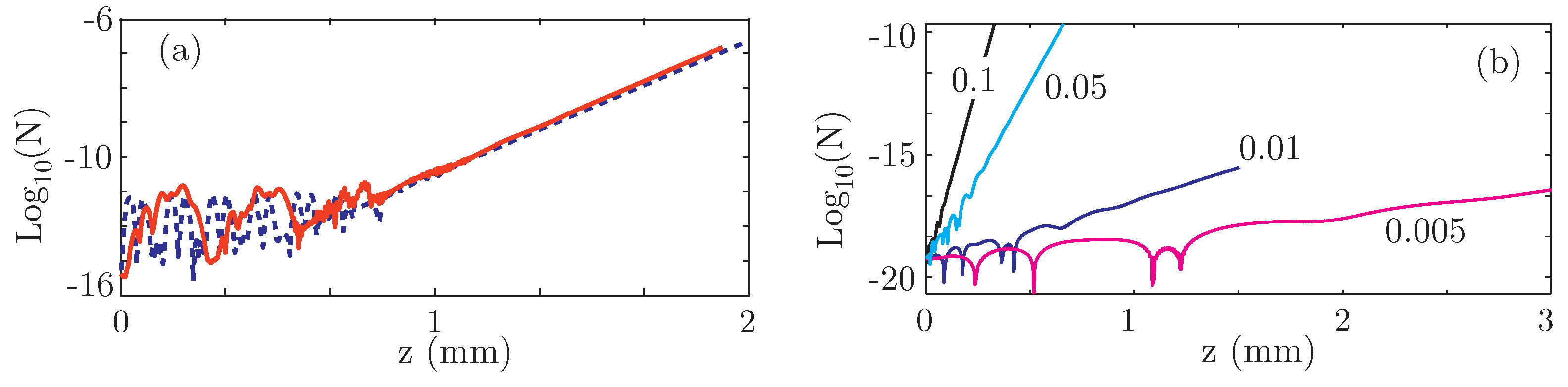}
\caption{(a) Evolution of the photon number N (in logarithmic scale) inside the black hole laser cavity for a $\delta n$ amplitude of 0.01 and two different configurations shown in Fig.~\ref{laser} (solid line - negative $\delta n$, dashed line - positive $\delta n$). (b) Photon number insied the laser cavity for various maximum $\delta n$ amplitudes, indicated in the graph. }
\label{long-laser}
\end{centering}
\end{figure}
\end{centering}

\subsection{Zero-frequency mode amplification}

Figures~\ref{laser-sim}(a) and (c) show results from a simulation in which the first several cavity round trips are shown. The horizon separation is 80 $\mu$m, with 2 $\mu$m input wavelength, $\Delta \zeta=3$ $\mu$m, $\delta n =0.01$ - the curves are normalised and displaced horizontally for viewing purposes.  The optical modes bounce back and forth but without any apparent overall increase in energy for nearly the first mm of propagation. At z=0.8 mm, a series of relatively weak bursts are clearly visible corresponding to T mode emission from the black hole horizon at each round  trip.
Figure~\ref{long-laser}(a) shows the total photon number inside the cavity. During the first $\sim1$ mm or propagation, the energy inside the cavity oscillates. These initial oscillations do not appear to be a numerical artefact and seem to correspond to real light pulse dynamics within the cavity. They most likely arise from a beating between the different modes that are concomitantly excited within the cavity and that therefore interfere with each other giving rise to amplitude oscillations, in a similar fashion to the build-up dynamics that may be observed in traditional laser cavities. After this short transitory period, a different regime appears in which the energy oscillations are washed out by a smooth and strong exponential increase of the total energy in the cavity. The origin of this change in behaviour can be understood from Fig.~\ref{laser-sim}(a).  During the first mm of evolution, light bounces back and forth but is gradually spread out to very low frequencies. This low frequency light has two effects: (i) due to its long wavelength it will start to completely fill the cavity and (ii) it will seed mode conversion to high frequency modes that lie close to or at the intersection of the dispersion curve with the horizontal axis, i.e. it will excite modes that have $\omega^{\prime}\rightarrow 0$, indicated with filled squares in Fig.~\ref{laser-sim}(b) (we name the high frequency intersection as the ``Z'' mode).  In order to understand why these zero-frequency modes  appear we recall that, as a consequence of the planckian distribution of Hawking emission, amplification at the horizon  diverges as $1/\omega^{\prime}$ for small $\omega^{\prime}$. Therefore, the laser gain experienced by these very low frequency modes is remarkably stronger with respect to all other modes and they will therefore quickly take over: high-gain lasing at the ``zero frequencies'' will, and is expected to dominate the black hole laser scenario.\\  %Evidence of the onset of this behaviour can also be seen in the last stages of the evolution shown in Fig.~\ref{short-laser}(d). \\
In Fig.~\ref{long-laser}(b) we show the cavity photon number for identical operating conditions and with varying maximum $\delta n$ amplitude, as indicated in the figure. The gain (slope of the curve) is seen to decrease with decreasing $\delta n$ amplitude as a result of the fact that we have fixed the distance over which the perturbation switches on, $\Delta\zeta=3$ $\mu$m. Therefore the horizon gradient (i.e. surface gravity) and hence also the horizon amplification decreases with decreasing $\delta n_{\textrm{max}}$. Nevertheless, significant laser gain is observed even at low perturbation amplitudes that hold promise for future experiments.\\%
\begin{centering}
\begin{figure}
\begin{centering}
\includegraphics[width=15cm]{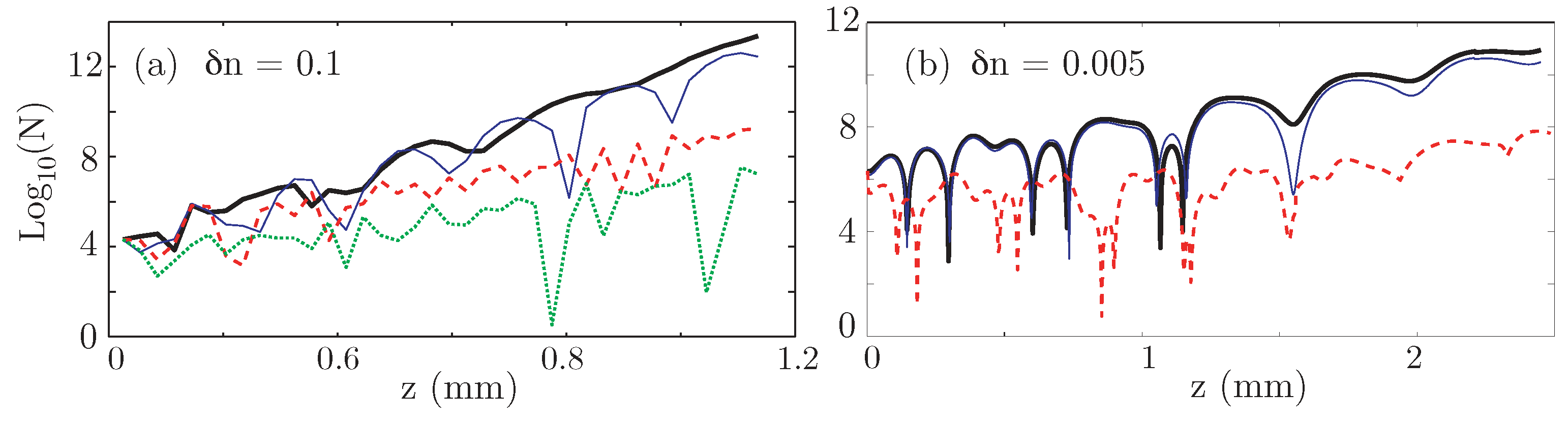}
\caption{Evolution of the photon number inside the cavity for (a) $\delta n=-0.1$ and (b) $\delta n=-0.005$. The simulations were performed by including a spectral filter that cuts all frequencies below a certain value. The shortest propagating wavelengths are (a) thick solid line - no filter, thin line - 20 $\mu$m, dashed line - 8 $\mu$m, dotted line - 4 $\mu$m. In (b) thick solid line - no filter, thin line - 60 $\mu$m, dashed line - 10 $\mu$m.}
\label{filtered}
\end{centering}
\end{figure}
\end{centering}
The question now arises as to how close to zero in the laboratory frame does the $\omega\rightarrow0$ zero-frequency mode  have to be in order for this process to occur? The question is relevant as real experiments cannot be expected to rely on modes for example in the radio frequency domain. We therefore repeated the simulations by including a low frequency filter so as to completely suppress all radiation below a certain threshold value. Fig.~\ref{filtered}(a) shows the results for a $\delta n$ amplitude of 0.1 and filters as indicated in the caption.
As may be seen, a filter placed at a wavelength of 20 $\mu$m hardly modified the overall gain and a reduction of $\sim$50\% is obtained only when the filter is placed at a wavelength of 8 $\mu$m. In Fig.~\ref{filtered}(b) we repeat the simulations for an experimentally realistic $\delta n$ amplitude of 0.005. A filter placed at 60 $\mu$m does not modify the laser gain and a 50\% cut is observed with a filter placed at 10 $\mu$m. This is extremely promising. Indeed, many media are transparent in to the  terahertz region (30-100 $\mu$m). For example diamond is transparent up to roughly 100 $\mu$m. This  would therefore seem to indicate that this kind of novel amplification process could be observed in real settings. \\

\section{Conclusions}
We have numerically analysed the possibility of observing black hole lasing using laser-pulse induced (or optical) horizons. The general behaviour of the black-white hole horizon cavity is very similar to the behaviour of similar cavities studied theoretically and numerically in Bose-Einstein condensates. Our numerical simulations have been carried out over longer propagation distances (i.e. larger number of cavity round-trips) with respect to previous studies and highlight the onset of a regime in which the cavity leads to a broadening of the mode spectra until the two comoving zero-frequency modes are excited (in the laboratory frame, one mode lies at zero-frequency, the other lies in the visible or ultraviolet region, depending on the specific dispersion relation of the medium). Due to the $1/\omega^{\prime}$ dependence of the horizon amplification, the zero-frequency modes oscillate in the cavity with a much higher gain.
These findings are summarised in the accompanying video animation that shows the evolution of the typical situation studied here: the initial optical pulse seeded in to the cavity is actually much shorter than the cavity itself (differently from typical cases studied so far in which the cavity length was of the same order of the input wavelength) and bounces back and forth until the zero-frequency modes take over and completely fills the cavity.\\
 Our numerical simulations treat the case of a coherently seeded laser. Naturally, one would expect that if the laser were to be seeded by vacuum fluctuations, then most certainly laser oscillation would occur at the zero-frequency modes as these have the highest gain and lowest lasing threshold. \\
A remarkable and unique feature of black hole lasers is the coupling between the very low frequency and very high frequency components of the electromagnetic spectrum. However, more complicated dispersion relations that include also resonances at certain wavelengths are expected to complicate this picture although a full model that includes  losses due to material absorption would be required in order to correctly model such a situation.\\
These simulations underline the fact that an experimental demonstration of black hole lasing, although not simple may not be as far-fetched as one may think. Refractive index perturbations of the order of 0.005 (or smaller) will give rise to significant gain over distances of the order of 1 cm. The dispersion relation used in this work is  very close to that of diamond, a material that is now widely used and can be shaped in to optical waveguides. Diamond exhibits not only a very simple dispersion curve but (as a consequence of this) also a remarkably wide transparency range that could easily sustain the zero-frequency amplification regime reported here. The laser cavity, i.e. the black and white hole horizons could be generated by focusing for example two independent and ultrashort laser pulses, e.g. at 800 nm wavelength: each pulse would create a $\delta n$ through the nonlinear Kerr effect and the relative delay between the two pulses would allow to control the laser cavity length. The main challenge would be to find a condition (laser wavelengths, durations, energies) such that the rising and falling edges of the laser pulses that form the horizons are maintained shorter than $\sim3$ $\mu$m for significantly long propagation distances, e.g. $>1-2$ mm. Future studies will hopefully unravel other settings or combinations of materials and wavelengths that may indeed lead to the first experimental black hole laser. In the meantime, this remains a fascinating scenario in which to combine technologies and ideas developed in the area of photonics to the study of flowing media and horizon physics.

\section{Acknowledgements}
The authors acknowledge financial support from EPSRC, grants EP/J00443X/1 - P/J004200X/1 and from the British Council, IAESTE student exchange program. 

\section*{References}
\newcommand{\noopsort}[1]{} \newcommand{\printfirst}[2]{#1}
  \newcommand{\singleletter}[1]{#1} \newcommand{\switchargs}[2]{#2#1}

\end{document}